\documentclass[%
twocolumn,
 amsmath,amssymb,
 aps,
prb,
]{revtex4-1}

\usepackage{graphicx}
\usepackage{dcolumn}
\usepackage{bm}
\usepackage{subcaption}
\usepackage{multirow}
\usepackage{color}

\begin{document}


\title{Topological and Thermoelectric Properties\\
of Double Antiperovskite Pnictides}

\author{Wen Fong Goh}
\author{Warren E. Pickett}%
\affiliation{%
 Department of Physics, University of California Davis 95616 CA, USA
}%

\date{\today}

\begin{abstract}
Doubling the perovskite cell (double perovskite) has been found to open new possibilities for engineering functional materials, magnetic materials in particular. This route should be applicable to the antiperovskite (aPV) class. In the pnictide based double aPV (2aPV) class introduced here magnetism is very rare, and we address them as new topological materials, possibly with thermoelectric interest. We have found that the 2aPV supercell provides a systematically larger band gap that can serve to inhibit bulk conductivity, and also large spin-orbit coupling (SOC) for band inversion.  We present examples from a broad study of double antiperovskites focusing on the X$_6$AA$'$B$_2$ configuration, where X is the alkaline earth element and A and B are the group 5A pnictogens. We find that an ``extended s'' state at the valence band minimum, described alternatively as a cation valence state or a modulated interstitial planewave state, plays a crucial role in both topological and thermoelectric properties. 
Several of these compounds may house topological phases, while transport calculations indicate they may also find themselves useful in thermoelectric applications.
\begin{description}
\item[PACS numbers]
\end{description}
\end{abstract}

\pacs{Valid PACS appear here}
\maketitle


\section{Introduction}

In spite of many existing classes of proposed topological insulators (TIs), the search for new classes with more favorable properties continues.  One important factor is that most existing TIs are defective in the interior, thus are not insulating enough in the bulk to allow study and potential application of their surface bands.  
One recipe for finding new TIs is to look for small gap insulators that have  valence and conduction bands with opposite parity and with small or negative gaps, and a different chemistry that might promote stoichiometry. For very small gap materials, the bands may be inverted by spin orbit-coupling (SOC), leading to insulating gaps that house topological states.\cite{Kane2005}   Ideally, the band gap (without SOC) must be small enough for band inversion, while large enough to inhibit bulk conductivity and enable application at room temperature. The other possibility is to have a band overlap semimetal (before SOC), with a gap opened at the Fermi level by SOC. In both cases the strength of SOC governs the magnitude of gap that can be obtained. This realization has focused attention on heavy atoms with large SOC.

Thermoelectric properties, based foremost on a large Seebeck coefficient, also in essence require small band gaps, since one wants a large derivative of the electronic density of states at the Fermi level, $dN(E)/dE|_{E_F}$, with strong particle-hole asymmetry and a low carrier density to minimize electronic thermal conduction. Unlike topological properties which depend sensitively to the momentum $\vec k$ dispersion of bands on either side of the gap, for thermoelectric properties the energy dependence [$N(E)$, and the square velocity $v^2(E)$ of carriers at energy $E$] is the focus. At this level of discussion any connection between topological and thermoelectric properties not evident, aside from the importance of a small or possibly zero gap.

Compounds containing heavy atoms become of special interest. First, if gapped, the gap often is small, favorable for both thermoelectric and topological properties. Of course SOC is large, increasing the likelihood of band inverson and topological character. However, a connection, though somewhat indirect, has been predicted and demonstrated by Singh and collaborators.\cite{singh,singh2} The key variable is the strength of SOC, which is large for heavy atoms, and the observation that in highly itinerant materials this strength can be varied quasi-continuously by isovalent doping. Substituting, say, Te with Se decreases the SOC strength, thereby tuning the band structure near the gap and adjusting both topological and thermoelectric properties. The predicted effect was verified by contrasting Bi$_2$Te$_3$ with isostructural and isovalent Bi$_2$Te$_2$Se.\cite{singh2} 

As just pointed out, heavy elements such as bismuth (Bi) and tellurium (Te) are favorable for TIs. A handful of materials, viz. Bi$_2$Se$_3$ family \cite{Zhang2009,Zhang2010,Yazyev2010} and Bi- and Te-based perovskites,\cite{Jin2013,Yang2015} have been discovered to possess non-trivial topological characteristics.
Perovskites, primarily oxide-based ones, are one of the more prevalent structures to be explored and designed, due to their cubic structure and multitude of members.
Their electronic structures often display a narrow gap with band minimum and maximum at high symmetry points, \cite{Sun2010,Jin2012,Hsieh2014} implying that these materials may have potential in thermoelectric applications. \cite{Bilal2015} Doubling the structures will affect the dispersion and have been suggested in certain cases to increase the topological insulator's bulk energy gap.\cite{Pi2017,Lee2017}

In this paper, we study the topological nature and the electronic structure based thermoelectric properties of several pnictide-based double antiperovskites, based on our previous study on single antiperovskite compounds \cite{Goh2018}. Our methods of calculations are described in Sec. II, followed in Sec. III by a description of the structures. In Sec. IV the calculated band gaps and inversion energies are provided for the $X_6AA'B_2$ class of compounds, where $X$ is one of the divalent alkaline earths Ca, Sr, and Ba; $AA'$ are pairs of heavy pnictides SbAs, BiAs,and BiSb; $B$ is one of the lighter pnictides N, P, and As.  The electronic structures of two of the compounds are illustrated in Sec. IV. In Sec. V the possibilities for topological properties are outlined, including effects of uniaxial strain. The thermoelectric coefficients of two illustrative compounds are presented in Sec. VI, and Sec. VII provides a brief summary.   

\section{Methods}

The electronic structure calculations are done with the full-potential
local orbital (FPLO) code \cite{Koepernik1999},
using the generalized gradient approximation (GGA) exchange-correlation
of Perdew, Burke, and Ernzerhof \cite{Perdew1996}.
A dense $20 \times 20 \times 20$ $k$-mesh was used for self-consistency because
of the delicate band overlap near $\Gamma$.
Spin-orbit coupling (SOC), in fact all relativistic effects, was included precisely by using the 
fully relativistic four component Kohn-Sham-Dirac equation 
implemented in FPLO, without resorting to the customary intermediate scalar relativistic approximation. 

The calculations of thermoelectric properties are done in BoltzTraP \cite{Madsen2006} by solving the Boltzmann equation via band interpolation scheme 
based on the band energy obtained from WIEN2k calculation \cite{Blaha2001}.
A dense k-point mesh of 125000 (3107 in IBZ) are used and interpolated onto a mesh of 20 times denser.
Bands within 4 eV around the Fermi level are used in the integration, with a fine energy mesh of 0.5 meV.

\section{Crystal Structure}

\begin{figure}
\centering
\includegraphics[width=0.3\textwidth]{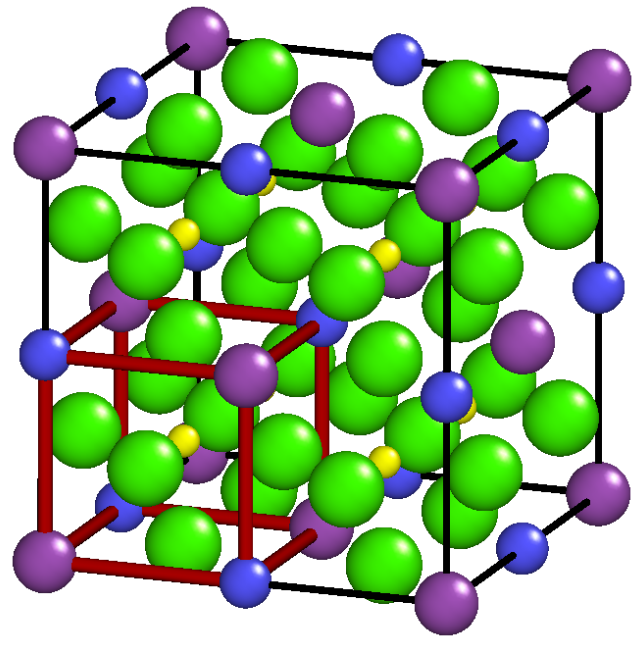}
\caption{Crystal structure of a double antiperovskite with $A$ (purple) $\neq$ $A'$ (blue) and $B = B'$ (small yellow spheres). Cations (green) form octahedra surrounding the $B$ anions.}
\label{crystal-doubleperov}
\end{figure}

The fcc double antiperovskite structure in Fig. \ref{crystal-doubleperov}
contains two alternating unit cells of a single antiperovskite cube, 
denoted in generality by $X_6AA'BB'$, where $X$ is an alkaline earth element 
and $A,A',B,B'$ are pnictides elements on the A and B sites of the 
perovskite structure. Since the $B$ and $B'$ anions are surrounded by 
$X_6$ octahedra, if $B\neq B'$ then $X$ is not required by symmetry to lie 
midway between them. However, $A\neq A'$ but $B=B'$, $X$ does still reside 
on the special site midway between $X$ ions. With distinct $A$-site atoms but 
$B=B'$, the spacegroup is $Fm\bar{3}m$ (\#225). Both $A$ and $A'$ have a 
site symmetry of $m\bar{3}m$, while $B$ and $X$ have a site symmetry of 
$\bar{4}3m$ and $mmm$ respectively.  The optimized lattice constants of 
the double aPVs are close to twice of that of the single aPVs.

We are interested in $A\neq A'$ and $B=B'$ because the $A$ site pnictide provides the upper valence bands at the gap (or band overlap). Thus modulation on the $A$ site is of particular interest.Based on our previous study on the topological characteristics of antiperovskites, 21 double antiperovskites $X_6AA'B_2$ involving heavy elements, where $X=$Ca, Sr, Ba, $A$ = Sb or Bi, $A'$ = Sb or As, and $B$ = N, P, and As, 
were selected for the study of potential topological characteristics. 

\section{Electronic Structure}

\begin{figure}
\centering
\begin{subfigure}[b]{4in}
\includegraphics[width=0.95\textwidth]{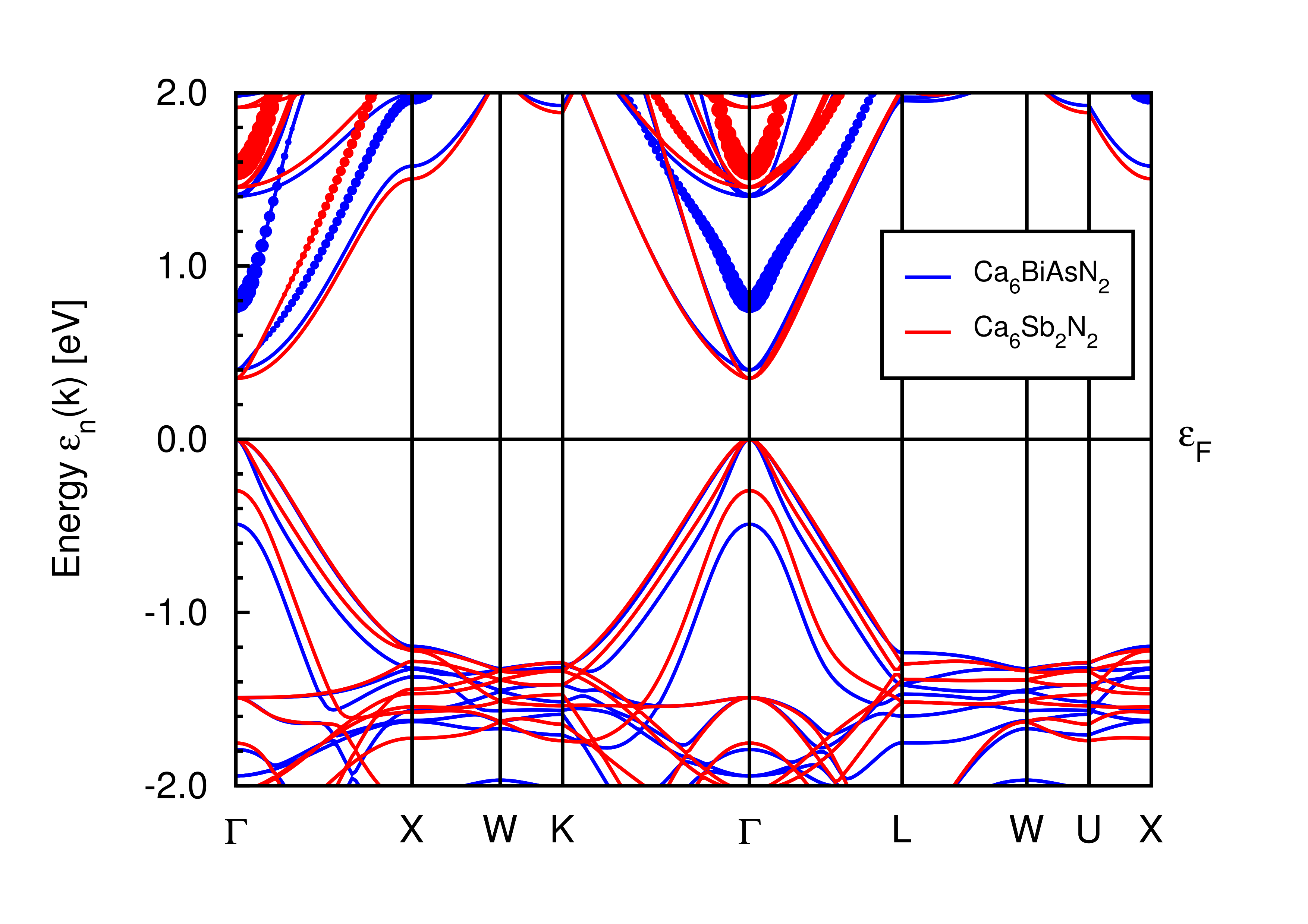}
\caption{$\mathrm{Ca_6BiAsN_2}$ vs $\mathrm{Ca_6Sb_2N_2}$.}
\label{n}
\end{subfigure}
\
\begin{subfigure}[b]{4in}
\includegraphics[width=0.95\textwidth]{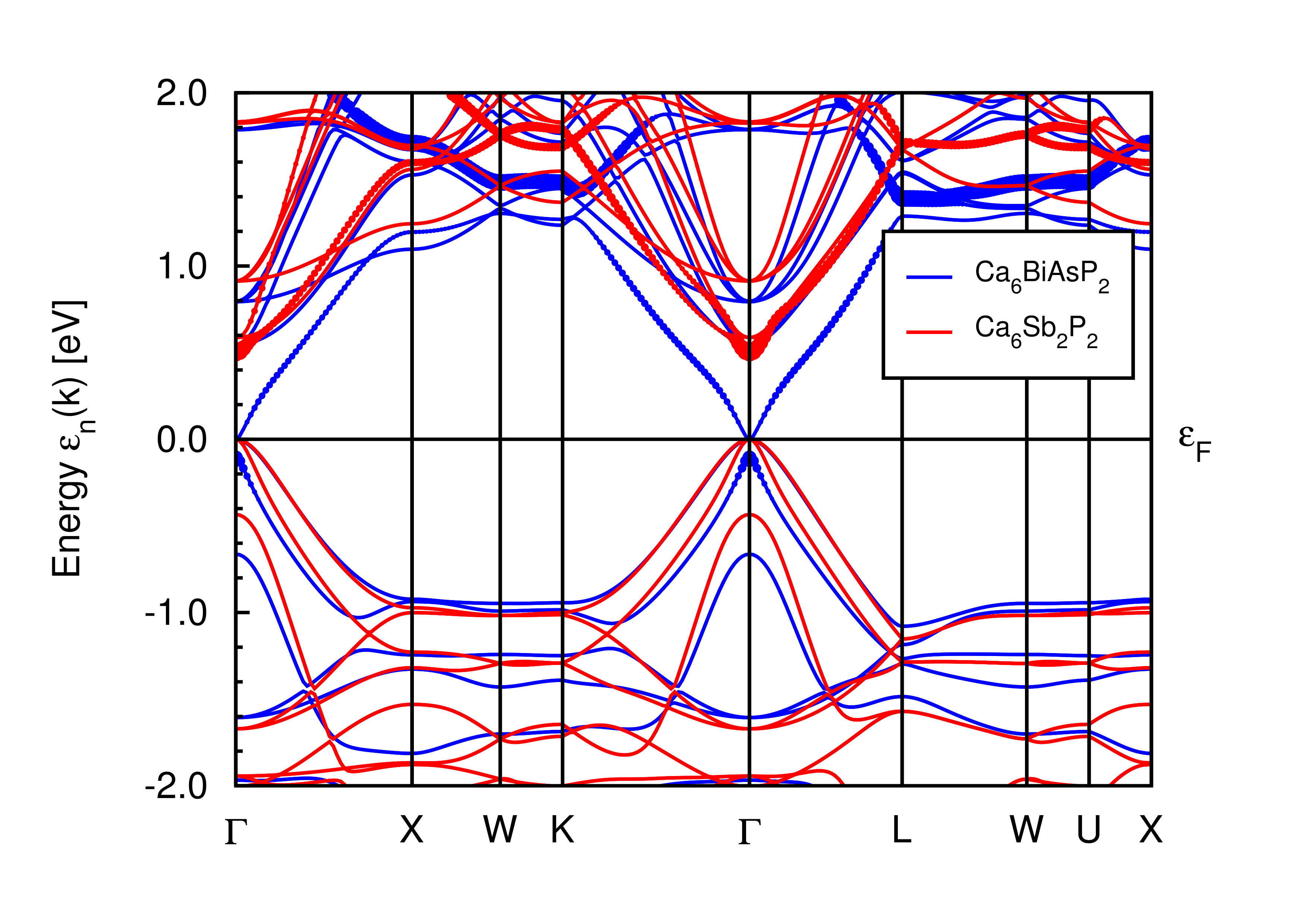}
\caption{$\mathrm{Ca_6BiAsP_2}$ vs $\mathrm{Ca_6Sb_2P_2}$.}
\label{p}
\end{subfigure}
\caption{Band structure of double antiperovskites, to compare and contrast $\mathrm{Ca_6BiAsN_2}$ and $\mathrm{Ca_6Sb_2N_2}$ (above), and $\mathrm{Ca_6BiAsP_2}$ and $\mathrm{Ca_6Sb_2P_2}$ (below). Spin-orbit coupling included in all plots.
The fat bands show even parity $s$-character of the heavy A atoms (As, Sb, Bi).}
\label{bandstruct-doubleperov}
\end{figure}

Near the gap (or band overlap) at $\Gamma$, which is the region of interest, these antiperovskites have valence bands contributed from the A site pnictide anion $p$ bands, while the conduction bands derive from the alkaline-earth cation $d$ bands.  Doubling to $B~-~B'$ pairs leaves the direct (positive or negative) gap at the $\Gamma$ point, with examples shown in Fig. \ref{bandstruct-doubleperov}. Studying the eigenvalues at the $\Gamma$ point then gives an idea of which candidates are most likely to be topological insulators.

As pointed out previously,\cite{Goh2018} the bands disperse quadratically from $\Gamma$ except for a single high velocity band which can become nearly linear if it approaches a valence band with small or negative gap, a peculiarity of this antiperovskite structure.
From the projected density of states (DOS) this unusual band has even parity $s$-character on both cation and anion sites, so we refer to this band as an extended $s$-like state (ext-$S$).
Since this ext-$S$ character has opposite parity to the occupied bands, inverting it with the odd parity valence band maximum state at the $\Gamma$ point transforms the system into a topological state.

A convenient way to determine if the system is in a topological insulating state is by looking at the band ordering at $\Gamma$, since band orderings at the other time reversal invariant momenta (high symmetry zone edge points) are widely gapped and never change in these systems. It is common, especially with heavy atoms, that SOC closes the band gap and inverts the band ordering. The band structure of $\mathrm{Ca_6BiAsP_2}$ with SOC shown in Fig. \ref{p} depicts a zero gap semiconducting state.  At $\Gamma$ the band at the Fermi level has a two-fold degeneracy (excluding spin degeneracy) and pushes the even parity ext-$S$-state into the occupied region. This band inversion leaves a topological zero-gap insulating state, with a $Z_2$ invariant of 1;(000).

It is instructive to compare the electronic structure of double antiperovskites to a reference single antiperovskites in double perovskite supercell structure, especially when the reference has a B atom that is midway between the two in the double aPV.
In Fig. \ref{bandstruct-doubleperov}, where SOC is included, the band gaps of $\mathrm{Ca_6BiAsN_2}$ and $\mathrm{Ca_6Sb_2N_2}$ are equal. Changing the cation to Sr gives a different result: $\mathrm{Sr_6BiAsN_2}$ is a zero gap semiconductor while  $\mathrm{Sr_6Sb_2N_2}$ has about 0.5 eV energy gap. The former is a topological zero gap semiconductor, the latter is a trivial semiconductor. The difference is the position of the ext-$S$ state at $\Gamma$.
The difference does however indicate that double aPVs are more likely to undergo band inversion than single aPVs.

\begin{table*}[]
\caption{This table provides the effect of mBJ gap correction to the GGA+SOC band energies, for the inversion energy (see text) and the band gap, of selected double antiperovskites. Results are presented for B site pnictogen smaller than the A, A' atoms, because these are the only energeticaly stable combinations. }
\label{table1-2}
\resizebox{\textwidth}{!}{
\begin{tabular}{|c|l|l|l|l|l|l|l|l|l|l|l|}
\hline
\multicolumn{2}{|c|}{$\mathrm{X_6AA'B_2}$} & B                & \multicolumn{3}{c|}{N}           & \multicolumn{3}{c|}{P}            & \multicolumn{3}{c|}{As}                 \\ \hline
AA'                    & X                   &                  & SOC     & SOC+mBJ & mBJ effect & SOC      & SOC+mBJ & mBJ effect & SOC        & SOC+mBJ   & mBJ effect   \\ \hline
\multirow{6}{*}{SbAs}  & \multirow{2}{*}{Ca} & Inversion Energy & 1.28 & 2.39   & 1.11    & 0.30  & 1.08   & 0.78    & \multicolumn{3}{l|}{\multirow{12}{*}{}} \\ \cline{3-9}
                       &                     & Band Gap         & 0.51 & 1.15   & 0.64    & 0.30  & 1.08   & 0.78    & \multicolumn{3}{l|}{}                   \\ \cline{2-9}
                       & \multirow{2}{*}{Sr} & Inversion Energy & 0.63 & 1.71   & 1.08    & 0.03  & 0.82   & 0.78    & \multicolumn{3}{l|}{}                   \\ \cline{3-9}
                       &                     & Band Gap         & 0.38 & 1.03   & 0.66    & 0.03  & 0.82   & 0.78    & \multicolumn{3}{l|}{}                   \\ \cline{2-9}
                       & \multirow{2}{*}{Ba} & Inversion Energy & 0.49 & 1.45   & 0.96    & 0.15  & 0.92   & 0.77    & \multicolumn{3}{l|}{}                   \\ \cline{3-9}
                       &                     & Band Gap         & 0    & 0.53   & 0.53    & 0.15  & 0.85   & 0.70    & \multicolumn{3}{l|}{}                   \\ \cline{1-9}
\multirow{6}{*}{BiAs}  & \multirow{2}{*}{Ca} & Inversion Energy & 0.82 & 1.92   & 1.10   & -0.07  & 0.69   & 0.76    & \multicolumn{3}{l|}{}                   \\ \cline{3-9}
                       &                     & Band Gap         & 0.40 & 1.05   & 0.65    & 0     & 0.69   & 0.69    & \multicolumn{3}{l|}{}                   \\ \cline{2-9}
                       & \multirow{2}{*}{Sr} & Inversion Energy & 0.29 & 1.36   & 1.07   & -0.29  & 0.48   & 0.76    & \multicolumn{3}{l|}{}                   \\ \cline{3-9}
                       &                     & Band Gap         & 0.29 & 1.00   & 0.70    & 0     & 0.48   & 0.48    & \multicolumn{3}{l|}{}                   \\ \cline{2-9}
                       & \multirow{2}{*}{Ba} & Inversion Energy & 0.31 & 1.29   & 0.97   & -0.12  & 0.63   & 0.75    & \multicolumn{3}{l|}{}                   \\ \cline{3-9}
                       &                     & Band Gap         & 0    & 0.56   & 0.56    & 0     & 0.63   & 0.63    & \multicolumn{3}{l|}{}                   \\ \hline
\multirow{6}{*}{BiSb}  & \multirow{2}{*}{Ca} & Inversion Energy & 1.07 & 2.09   & 1.02    & 0.10  & 0.84   & 0.74   & -0.17   	  & 0.56     & 0.73      \\ \cline{3-12} 
                       &                     & Band Gap         & 0.21 & 0.81   & 0.61    & 0.10  & 0.84   & 0.74    & 0          & 0.56     & 0.56      \\ \cline{2-12} 
                       & \multirow{2}{*}{Sr} & Inversion Energy & 0.51 & 1.52   & 1.01   & -0.14  & 0.62   & 0.76   & -0.35       & 0.39     & 0.74      \\ \cline{3-12} 
                       &                     & Band Gap         & 0.18 & 0.82   & 0.64    & 0     & 0.62   & 0.62    & 0          & 0.39     & 0.39      \\ \cline{2-12} 
                       & \multirow{2}{*}{Ba} & Inversion Energy & 0.53 & 1.43   & 0.90    & 0     & 0.75   & 0.74   & -0.18       & 0.55     & 0.73      \\ \cline{3-12} 
                       &                     & Band Gap         & 0    & 0.46   & 0.46    & 0     & 0.72   & 0.72    & 0          & 0.55     & 0.554      \\ \hline
\end{tabular}
}
\end{table*}

We have surveyed the isovalent class of Ae-Pn double aPV compounds.
Changing the $B$ atom to a heavier one, for example, from $\mathrm{Ca_6BiAsN_2}$ to $\mathrm{Ca_6BiAsP_2}$ lowers the position of the ext-$S$ band, giving smaller inversion energy and more likely topological character (see Table \ref{table1-2}). Similar conclusions result when changing the size of $A$ and $A'$ atoms, although sometimes the volume change affects this trend.
In summary, heavy elements favor band inversion but volume effects are relevant.
There are  cases for which the band inversion can arise without the help of SOC.
For example, in $\mathrm{Sr_6BiAsP_2}$ where the interplay of element size and volume of the system has reached the ideal point, band inversion occurs without SOC.


\section{Topological States}

The inversion energy is defined as the eigenvalue of the even ext-$S$-state minus that of the odd valence band p-state,
thus negative inversion energies will promote a non-trivial $Z_2$ index. From Table \ref{table1-2}, $\mathrm{\{Ca,Sr,Ba\}_6BiAsP_2}$, $\mathrm{Sr_6BiSbP_2}$ and $\mathrm{\{Ca,Sr,Ba\}_6BiSbAs_2}$ show
negative inversion energy. Since band calculations with GGA exchange-correlation often underestimate the experimental band gap and give spurious inversion results, it is common to include the modified Becke-Johnson (mBJ) exchange-correlation potential \cite{Becke2006,Tran2007,Tran2009}, which provides a self-energy-like correction to the eigenvalues.  On average, the GGA inversion energy is in the range of 0.3 eV while mBJ splits the valence and conduction bands by about 0.7 eV. This increase in gap is detrimental in obtaining topological states.

\begin{figure}
\centering
\begin{subfigure}[b]{4in}
\includegraphics[width=\textwidth]{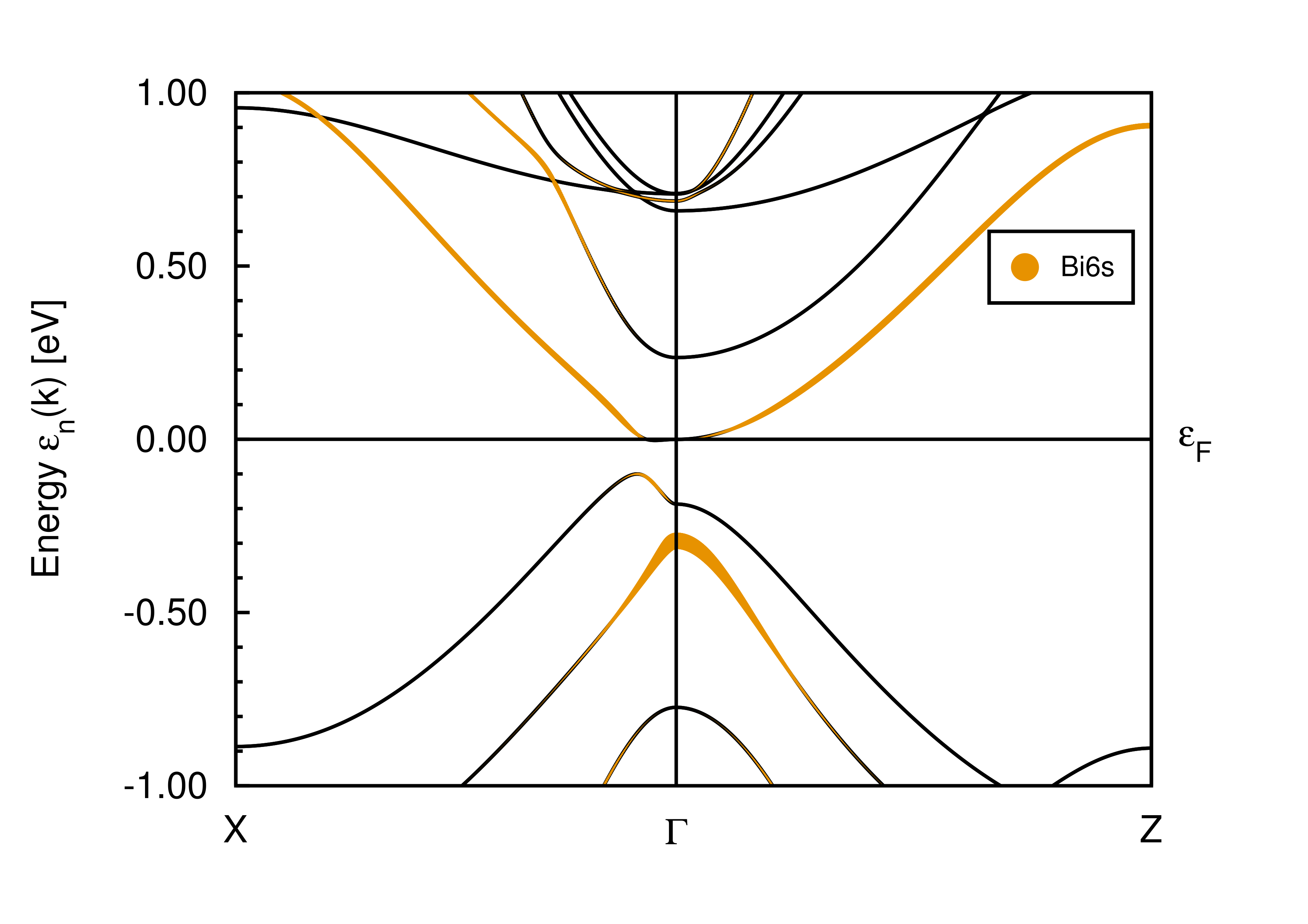}
\caption{}
\end{subfigure}
\
\begin{subfigure}[b]{4in}
\includegraphics[width=\textwidth]{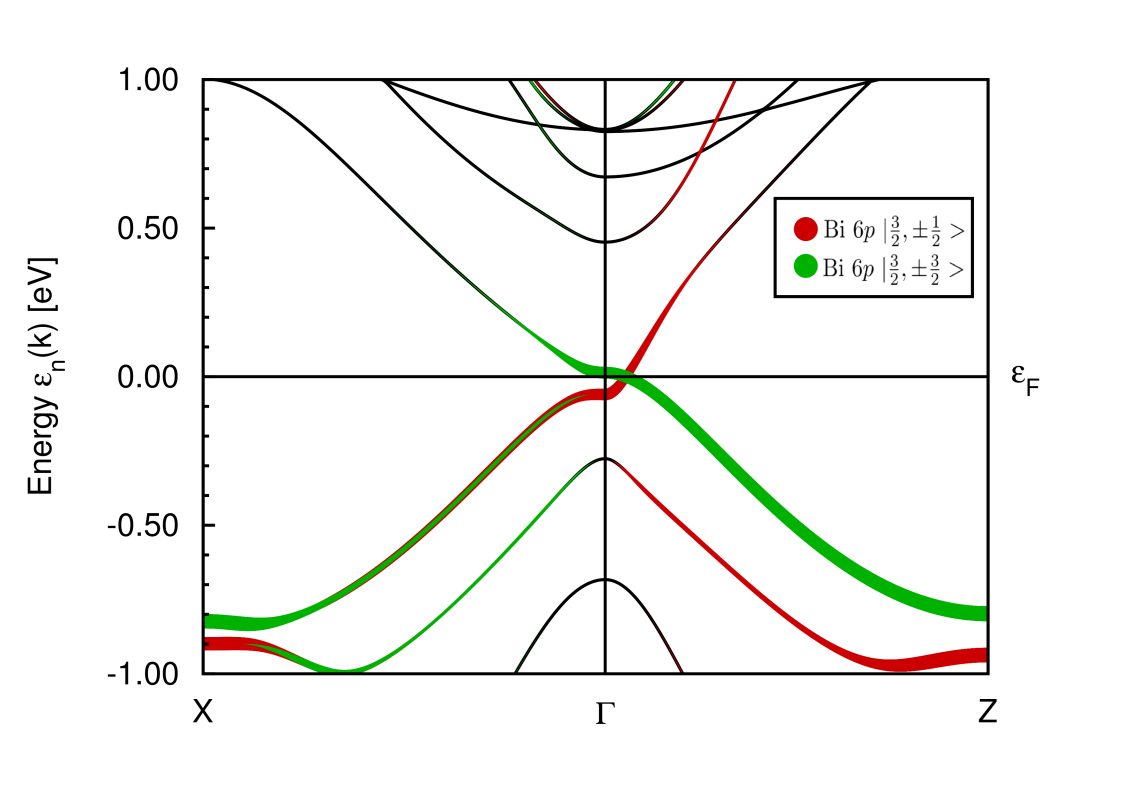}
\caption{}
\end{subfigure}
\caption{GGA+mBJ band structure of uniaxial (001) strained $\mathrm{Ca_6BiAsP_2}$, resulting in a topological insulator for 5\% compressive strain (top panel) or Dirac semimetal for 5\% tensile strain (bottom panel).}
\label{ti-dirac-doubleperov}
\end{figure}

However, strain can open up a gap by symmetry-allowed hybridization, while promoting the inverted band ordering with non-trivial topological invariant. Without SOC, for example, (001) strain splits the $p$ triplet at $\Gamma$ into a singlet and a doublet, with one {\it rising} in energy (narrowing the gap or inverting the bands) and the other being lowered, depending on the sign of strain. Similarly, SOC splits the doublet, with the eigenvalue that is displaced upward tending to decrease or invert the gap.  For  $\mathrm{Ca_6BiAsP_2}$ for example, the maximum gap achieved is 70 meV with compressive strain of $6\%$, but 110 meV in $\mathrm{Sr_6BiAsP_2}$ with compressive strain of $8\%$.
Fig. \ref{ti-dirac-doubleperov} shows the band structures of $5\%$ compressive and tensile strain.

Both signs of strain result in topological states.
Due to the band inversion at the $\Gamma$ point for the compressive case, the $Z_2$ invariant is [1;000], a strong topological insulator state results. The topological surface bands crossing within the bulk band gap will give rise to a Dirac point. Tensile strain, on the other hand, produces a Dirac semimetal, with bands gapped everywhere except near $\Gamma$ along $\Gamma$-$Z$, thus two symmetry related Dirac points. The dispersion around this Dirac point is highly anisotropic: the velocity toward $\Gamma$ is extremely small.

Fig. \ref{ti-dirac-doubleperov}(b) shows the corresponding band structures for these strains.  For $5\%$ tensile strained $\mathrm{Ca_6BiAsP_2}$ along $\Gamma$ - Z, the two double spin-degenerate bands with $j_z = \frac{1}{2}$ and $j_z = \frac{3}{2}$ cross, producing a four-fold degenerate Dirac point which, as mentioned, is also orbitally degenerate due to the mirror symmetry with respect to the $x-y$ plane. Strain therefore interpolates between surface and bulk topological behavior in Ca$_6$BiAsP$_2$.
The nearly flat band emerging from this Dirac point toward $\Gamma$ may have implications for the transport properties.

Table \ref{table1-2} shows the band gap and inversion energy, including SOC, of selected double aPVs with and without the mBJ gap correction.  Note that the inversion energy is defined as the energy difference between the``ext-S'' and the valence band maximum. Some compounds show negative inversion energy; the band orderings have been inverted by SOC. These inversion energies are at least half the value of that added to the gap by the mBJ shift, in which case the band ordering is no longer inverted. The band gaps and the inversion energies show a range of values; the mBJ shifts are consistent across the $B$ = (N, P, As) column. In general, compounds with nitrogen have a larger mBJ shift than the compounds with P or As. 


\section{Thermoelectric Properties}

\subsection{Formalism}
The semiconducting band structures in these compounds suggest possible candidates for thermoelectric applications, with the possibly of learning more about the connection of thermoelectric properties to electronic structure.\cite{singh}  Unlike the topological properties just discussed, which depend critically on positions in energy and dispersions in certain regions of the zone, the thermoelectric functions, viz. Seebeck coefficient $S$, 
power factor $P$, and figure of merit $ZT$ depend almost entirely on the distribution of available states -- the density of states N(E). As commonly done, we treat the elastic scattering time $\tau$ as energy independent and isotropic, the latter being particularly good in cubic materials such as those we have treated here. For several properties $\tau$ tne cancels out. 

The relations necessary to identify the origin of the calculated behavior are the following.
\begin{eqnarray}
S(T)&=& \nu(T)/\sigma(T),\nonumber \\
\frac{\sigma(E)}{\tau}&=&\frac{e^2}{3} N(E)v^2(E) \nonumber \\
\frac{\sigma(T)}{\tau}&=& \int dE \frac{\sigma(E)}{\tau}       [-\frac{df(E-\mu;T)}{dE}] \nonumber \\
\frac{\nu(T)}{\tau} &=&\frac{1}{eT} \int dE (E-\mu)\frac{\sigma(E,T)}{\tau}    [-\frac{df(E-\mu;T)}{dE}] \nonumber \\
\frac{\kappa_e(T)}{\tau}&=&\frac{1}{e^2T} \int dE (E-\mu)^2\frac{\sigma(E,T)}{\tau}    [-\frac{df(E-\mu;T)}{dE}] \nonumber \\
P(T)&=&S(T)\sigma(T)/\tau \nonumber \\
Z(T)T&=&S^2(T)\frac{\sigma(T)/\tau}{\kappa_e/\tau}.
\end{eqnarray}
$\kappa_e$ is the electronic contribution to the thermal conductivity. The actual figure of merit $Z$ includes a lattice contribution in the above equation as well as the electron one $\kappa_e$, so $ZT$ as given above, and displayed below, is an `electronic figure of merit.' $v^2(E)$ is the mean squared velocity averaged over the surfaces of constant energy $E$, and $f(E-\mu;T)$ is the Fermi-Dirac thermal distribution function. The Seebeck coefficient $S$ is the ratio of the thermal and electric conductivity response functions, $\nu$ and $\sigma$  respectively. The chemical potential $\mu(T)$ is determined at each temperature $T$ to conserve particle number: ${\cal N}=\int N(E)f(E-\mu)dE$. Under the stated conditions, the quantities above are independent of $\tau$. Note that the integrands in $\sigma$, $\nu$, and $\kappa_e$ involve increasingly higher powers of the carrier energy $E$ with respect to $\mu$, causing them to reflect more detail of the dispersion around the gap.

 We caution that the lattice thermal conductivity is not included in our results. It will lower the value of the total $ZT$, but realistic calculations in line with these electronic contributions are challenging, and require additional algorithms and codes. The options, as outlined by Stackhouse and Stixrude,\cite{Stixrude2010} are: the Green-Kubo thermal Green’s function method, which at its most basic level a fully quantum-mechanical approach; non-equilibrium first principles molecular dynamics; transient non-equilibrium molecular dynamics, a version of the method just above requiring additional algorithms. Each of these requires special codes and is far beyond the scope of this paper and of most works evaluating the electronic contribution to ZT. Differences between oxides (where more work has been done) and nitrides will be mentioned in the Summary.

Another point is that the lattice thermal conductivity is dominated, especially at the lower temperature end, by acoustic modes. The compounds we propose have very heavy atoms (Bi or Sb) that will give acoustic modes low group velocities. The velocity enters the thermal conductivity as the square (see Stackhouse and Stixrude\cite{Stixrude2010}), hence as 1/M (mass). In addition, these heavy atoms reside in the large A site of the perovskite cell and may be expected to “rattle” (have large amplitude and be strongly anharmonic), leading to a short relaxation time that will further reduce the lattice thermal conductivity. The correction to our `electronic figure of merit' thus may be expected to be small.

\begin{figure*}
\centering
\begin{subfigure}[b]{3in}
\includegraphics[width=\textwidth]{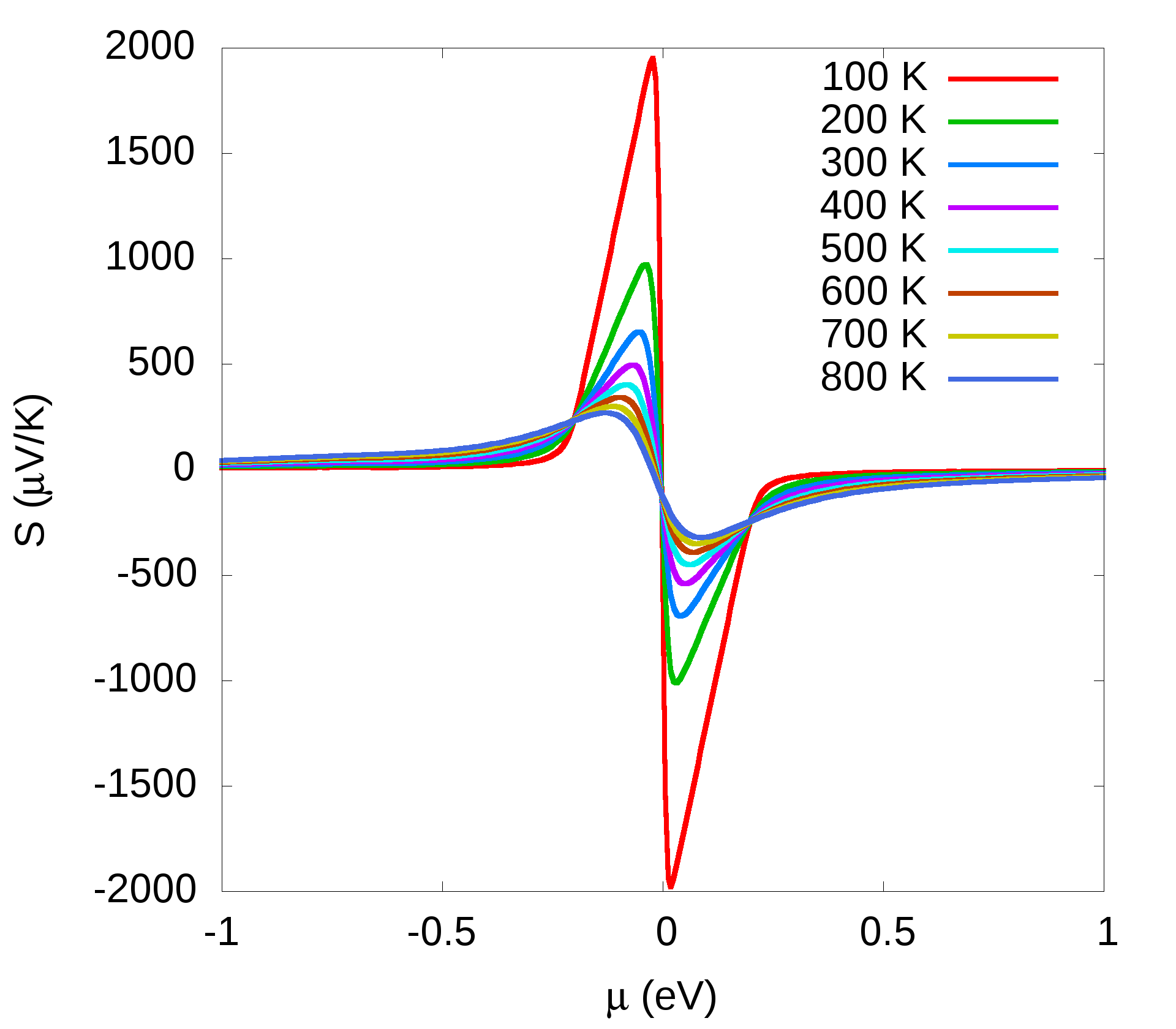}
\caption{}
\end{subfigure}
\begin{subfigure}[b]{3in}
\includegraphics[width=\textwidth]{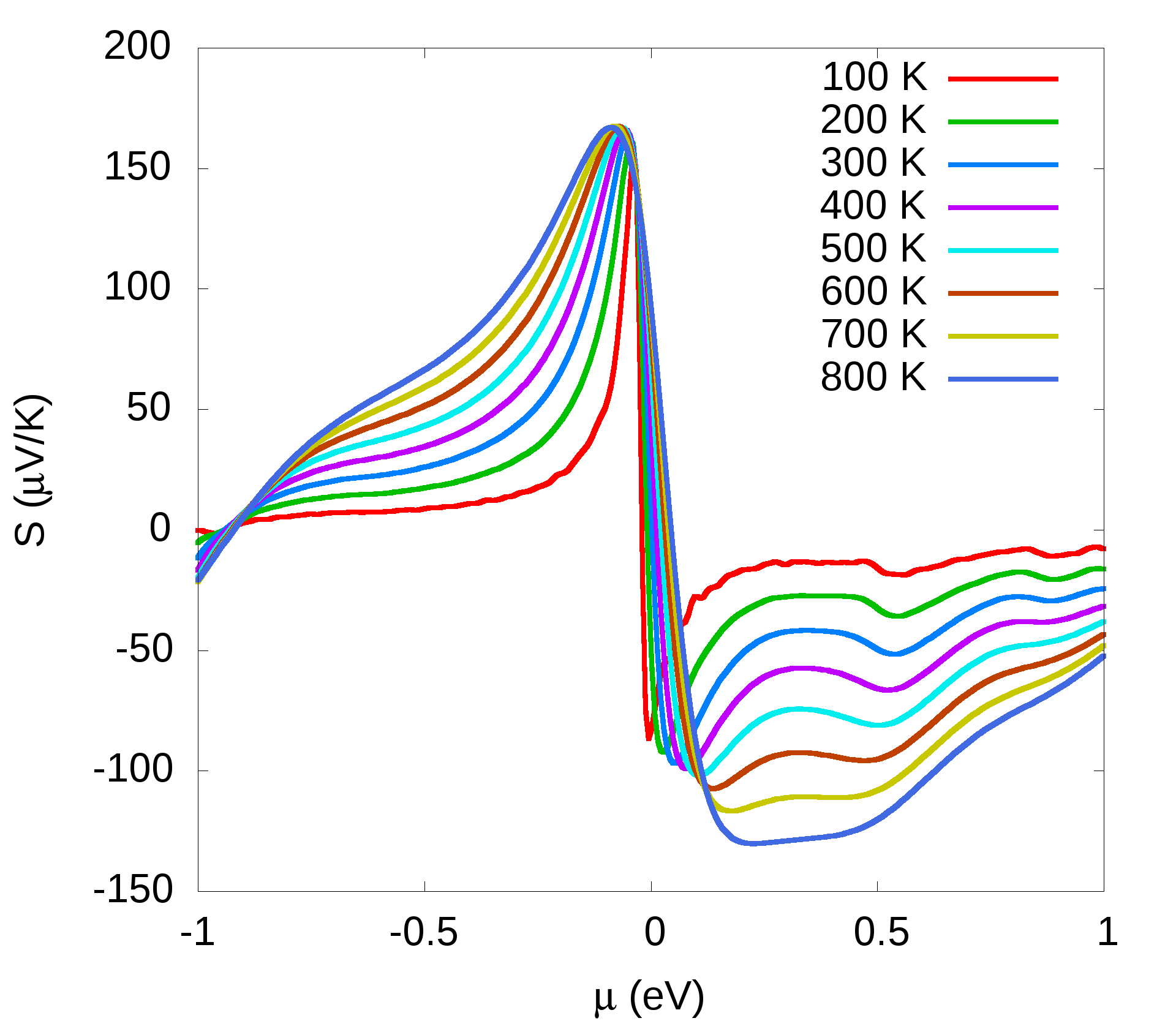}
\caption{}
\end{subfigure}
\
\begin{subfigure}[b]{3in}
\includegraphics[width=\textwidth]{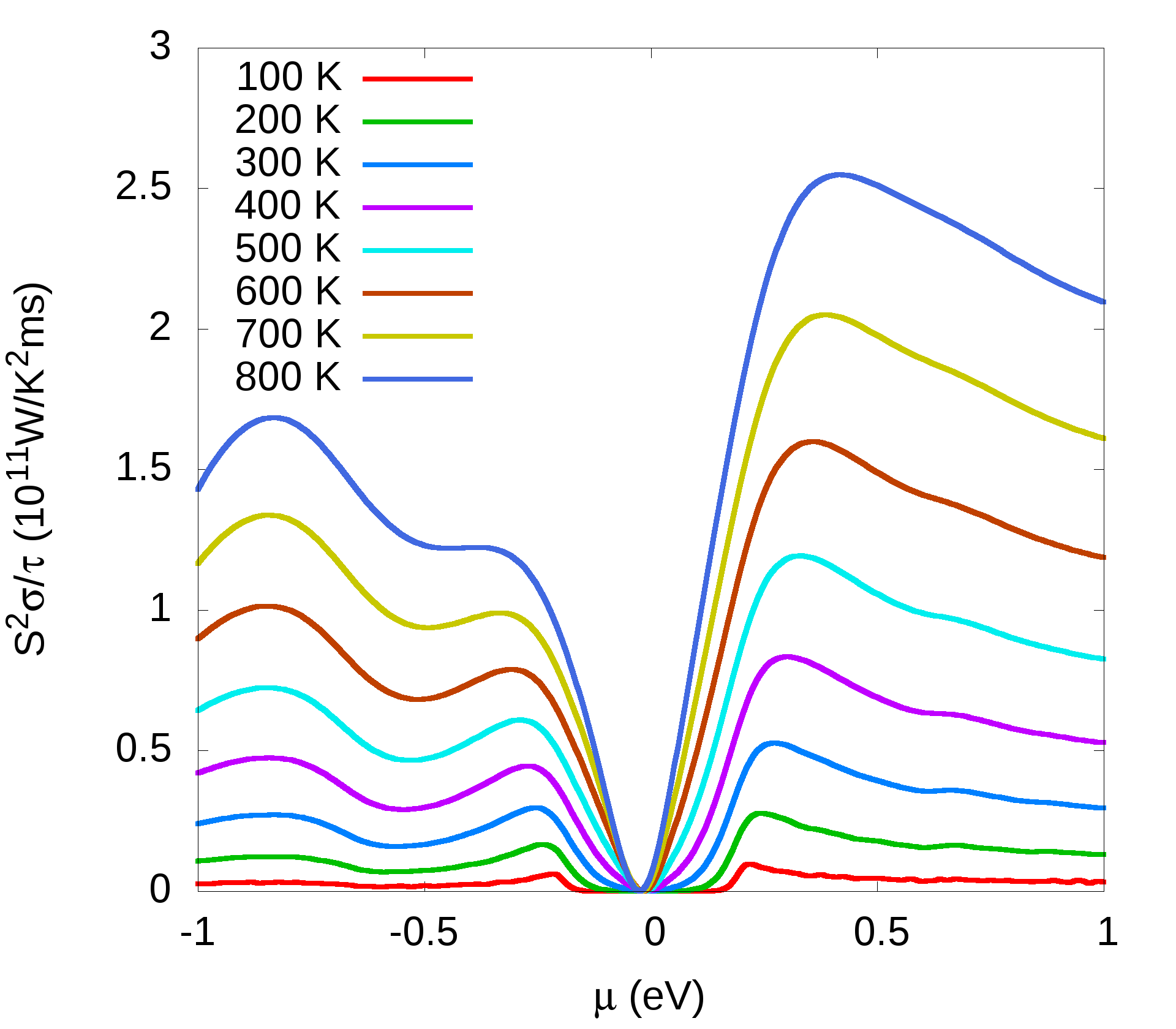}
\caption{}
\end{subfigure}
\begin{subfigure}[b]{3in}
\includegraphics[width=\textwidth]{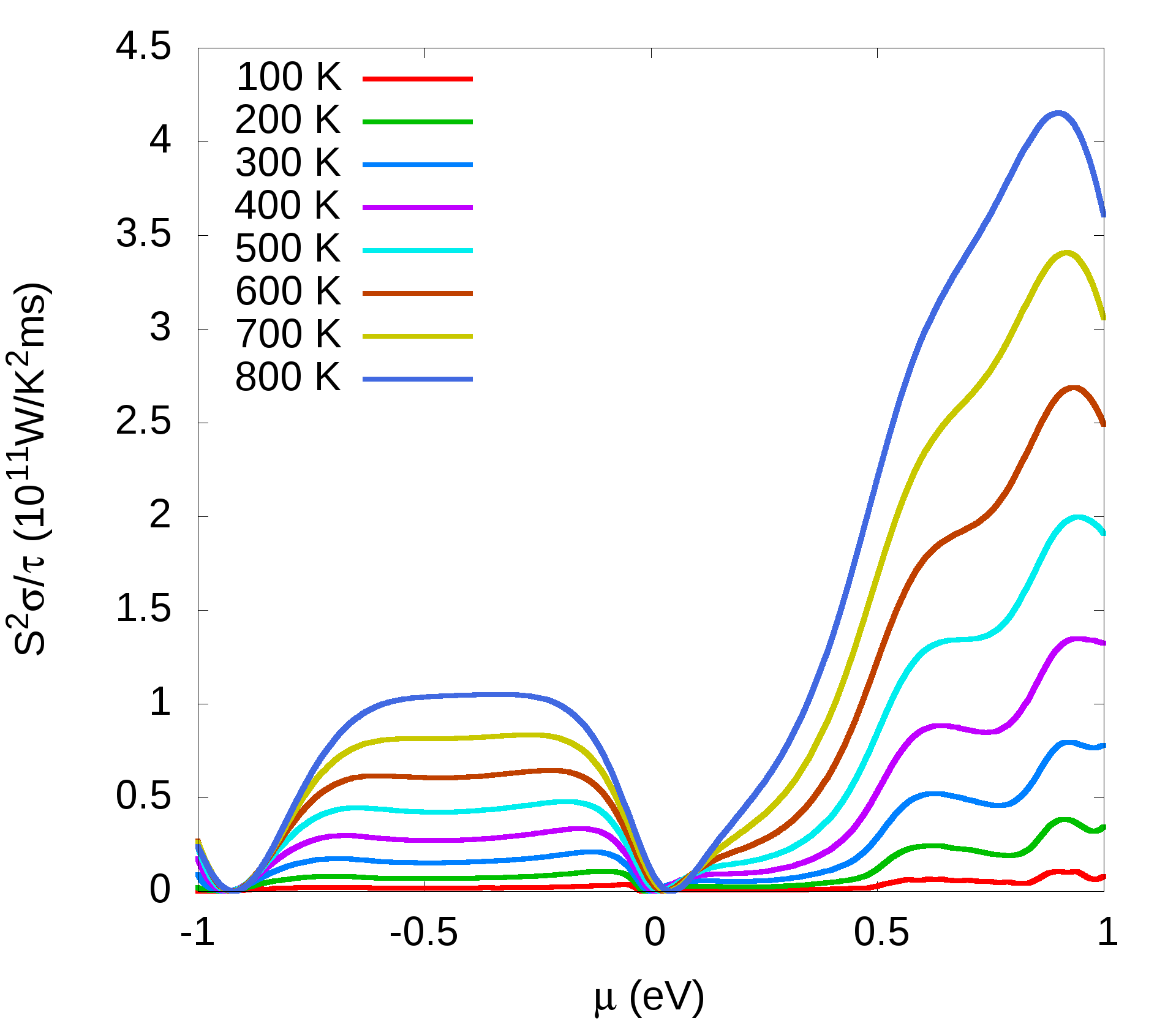}
\caption{}
\end{subfigure}
\
\begin{subfigure}[b]{3in}
\includegraphics[width=\textwidth]{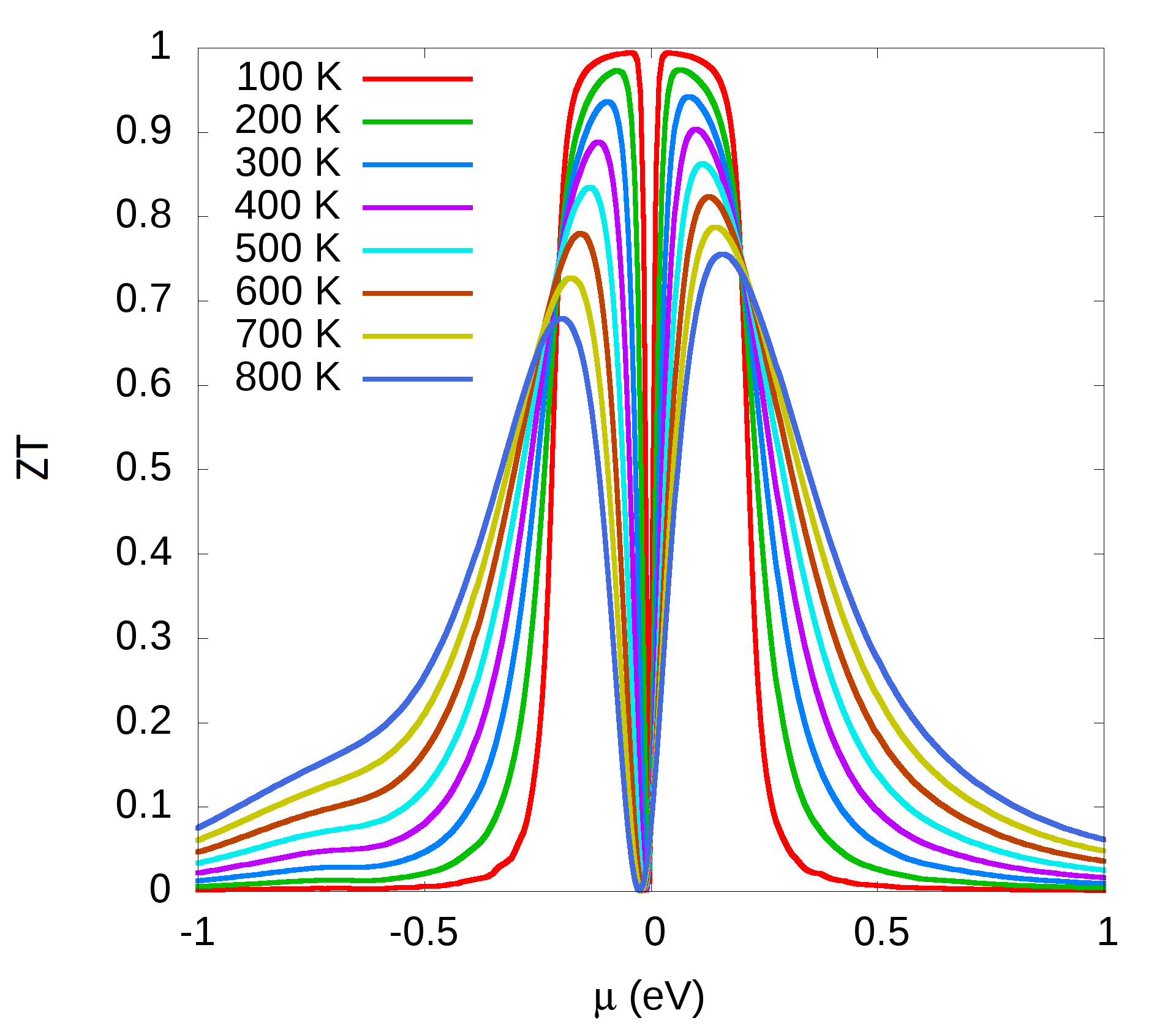}
\caption{}
\end{subfigure}
\begin{subfigure}[b]{3in}
\includegraphics[width=\textwidth]{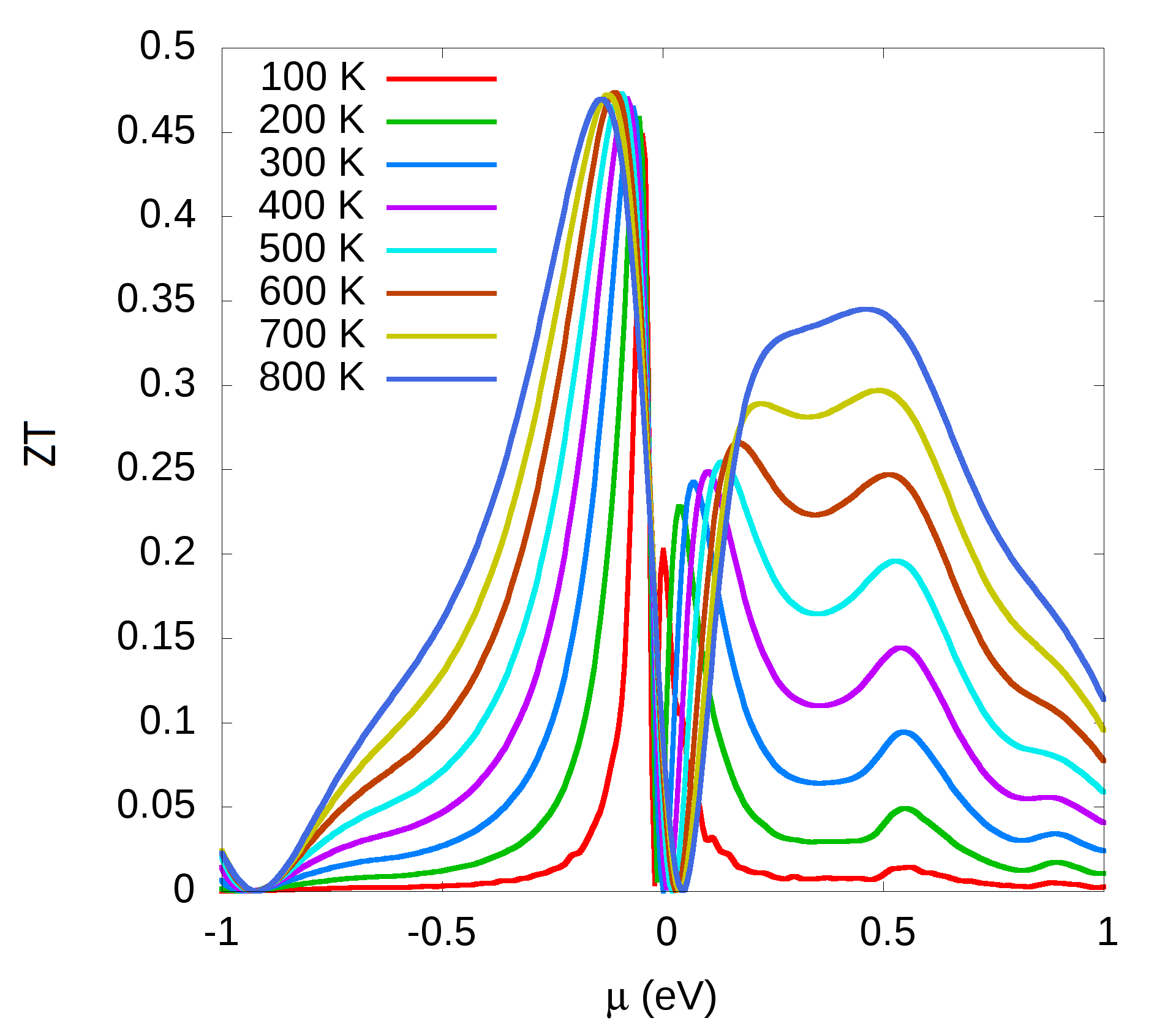}
\caption{}
\end{subfigure}
\caption{The Seebeck coefficient (top panels), thermopower (middle panels), and figure of merit $ZT$ (lower panels) of $\mathrm{Ca_6BiAsN_2}$ (left column) with small gap and $\mathrm{Ca_6BiAsP_2}$ (right column) with zero gap, versus chemical potential,  in the temperature range  $100\mathrm{K}-800\mathrm{K}$.
}
\label{thermo}
\end{figure*}

The examples of the electronic thermoelectric behavior with temperature versus chemical potential $\mu$ ({\it i.e.} doping level) are plotted in Fig. \ref{thermo} and will be discussed below. The ``independent variable'' $\mu$ in these plots account for both temperature and coping dependence. T-dependence is determined by the particle conservation mentioned just above. Doping dependence is treated in the rigid band approximation, which is reliable in itinerant systems in the low carrier density regime (where changes in band structure due to carriers is negligible). One then has the commonly quoted result that in the low temperature limit $S(T)\propto dN(E)/dE|_{E=\mu(0)}T$. Thus having $\mu(0)$ near a band edge or near other sharp structure in $N(E)$ leads to large values of $S(T)/T$.

\subsection{Examples}
Our discussion now will focus on two representative cases, the small gap (0.4 eV) semiconducting compound Ca$_6$BiAsN$_2$ (CBAN) and zero gap semiconductor Ca$_6$BiAsP$_2$ (CBAP).  For both types of spectra,
$\kappa_e(\mu)$ is not shown, as its behavior is simple: it has a minimum at the intrinsic chemical potential and rises quadratically to rather high doping levels, and the temperature dependence is unremarkable. The commonly studied electronic conductivity $\sigma$ (also not shown) is a thermally broadened version of transport product $N(E)v^2(E)$ which is a much more slowly varying function of $E$ than $N(E)$ and $v^2(E)$ separately. 

The Seebeck coefficient shows the usual (for low doping) positive sign for hole doping $\mu<0$ and negative sign for electron doping $\mu>0$. The calculated value reaches nearly 2000 $\mu$V/K at 100 K for CBAN, dropping to around 700 $\mu$V/K at room temperature. In the more relevant coefficients, this high value is tempered by other energy-dependent factors. Unfortunately, the thermopower $P(T)$ does not make good use of the large values of $S(T)$ very near the gap, because the conductivity is low there. The theromopower becomes very low for $\mu$ within the gap, due to the extremely low conductivity there. Unlike the Seebeck coefficient which is nearly symmetric between electrons and holes, the thermopower reveals asymmetry, being nearly 40\% lower for the holes at comparable doping levels.

For zero gapped CBAP, the Seebeck coefficient is an order of magnitude smaller than for CBAN, and the particle hole asymmetry is evident. This asymmetry of the effective masses is visible in Fig.~2. The conduction band minimum of CBAP at the $\Gamma$ point has very small mass followed by higher velocities, due to the steep dip in conduction band at $\Gamma$ also evident in Fig. 2. Since its valence band maximum remains free-electron-like, the Seebeck coefficient is asymmetric and somewhat shifted towards negative chemical potential. The calculated electric conductivity and electronic thermal conductivity over relaxation time (not shown) of CBAN and CBAP are similar; they are further from the chemical potential and larger at higher temperature due to the enhancement of carrier concentration.

The thermopower and figure of merit $ZT$ quantify the potential efficiency in thermoelectric applications. The power factor of CBAN shows two peaks around $\mu=0$, while that of CBAP shows a higher peak at positive chemical potential, indicating that better thermoelectric properties can be realized by electron doping. With rising temperature, the value of the power increases by five times and eight times for CBAN and CBAP respectively, from room temperature to 800 K.

At 100 K, $ZT$ of CBAN is calculated to be unity for low doping levels of either sign. This dimensionless thermoelectic device efficiency decreases slowly with rising temperature, dropping by only 5$\%$ at room temperature. This small decrease indicates that this material has the potential to produce thermoelectic power effectively at low temperature.
For zero gap CBAP, the value is half that of the small gap semiconductor. Nevertheless, these values are among the highest found in the literature,\cite{Li2017,Bhamu2017}
including these promising thermoelectric double perovskites \cite{Aguirre2019,Sahnoun2017,Takahashi2012,Roy2016,Arribi2016}.
Our results indicate that these pnictide-based double antiperovskites have the potential to be competitive thermoelectric materials.

\section{Summary}

The topological characteristics and thermoelectric properties of representative pnictide-based double antiperovskites have been studied.
Doubling the perovskite structure offers a larger bulk insulating gap than single perovskite provides, plus they are more likely to be inverted by SOC.
Based on the GGA exchange-correlation functional,
uniaxial compressive strain opens up an energy gap at the $\Gamma$ point producing a topological insulator with a large bulk band gap.
Tensile strain along $z$ direction, on the other hand, gives a Dirac semimetal with Dirac band crossing along $\Gamma$ - Z.
However, an uncertainty remains whether these topological materials can be realized, as the corrections to the band energy calculation with mBJ potential suggest trivial insulators.

Nonetheless, they are potential candidates for thermoelectric applications due to the high values of Seebeck coefficient, thermoelectric power and figure of merit.

\section{Acknowledgments} W.F.G. was supported by NSF Grant DMR 1534719, while W.E.P. had support from DOE grant DE-FG02-04ER46111.

\end{document}